\documentclass[aip,jcp,reprint]{revtex4-1}

\usepackage{graphicx}
\usepackage{dcolumn}
\usepackage{bm}
\usepackage{hyperref}
\usepackage[mathlines]{lineno}
\usepackage{color}
\usepackage[normalem]{ulem}
\usepackage{rotating}
\def\he4{$^4$He}
\begin{document}


\title{Superfluidity of quasi-2D $^4$He droplets on graphite}

\author{K. M. Kolevski}
\author {M. Boninsegni}
\email{m.boninsegni@ualberta.ca}
\affiliation{Department of Physics, University of Alberta, Edmonton, Alberta, Canada T6G 2H5}

\date{\today}

\begin{abstract}
The superfluid response of nanoscale size quasi-2D \he4 droplets adsorbed on a graphite substrate is investigated by computer  simulations. It is found that clusters comprising as few as 7 atoms are stable at temperatures of $\lesssim$ 0.15 K. Clusters of $\sim$ 20 atoms or less are liquid-like and  $\sim$100\% superfluid. As the size is increased, the central region crystallizes, forming the commensurate $C_{1/3}$ phase, with a quasi-1D surface layer of superfluidity possibly evolving into a Luttinger liquid or a transverse quantum superfluid with increasing cluster size. The relevance to quasi-2D molecular spectroscopy is discussed.

\end{abstract}
\maketitle
The investigation of small, highly quantal clusters, is actively pursued to gain insight into the emergence of bulk properties such as superfluidity (SF) or crystalline order, and their interplay as the size of the system is increased. For example, significant progress has been afforded in the understanding of the microscopic origin of SF through the extensive theoretical and experimental investigation of \he4 droplets \cite{Sindzingre1989,Toennies2001,Toennies2022}.
\\ \indent
It seems compelling now to extend this investigation to the (quasi) two-dimensional case. This is justified by the significant body of research on helium films, spanning now several decades, which has revealed unique quantum behaviors and novel physical phenomena that emerge in reduced dimensions \cite{Reppy1980,Taborek2020,Hallock2021}, and that could similarly manifest in 2D helium clusters. Generally speaking, fundamental issues such as the smallest size of a superfluid cluster, or the interplay of superfluidity and atomic localization, may be addressed more conveniently and directly in a quasi-2D setting than in three dimensions, given the modern, powerful 
substrate atomic imaging technologies that are available (see, for instance, Ref. \cite{Shibata2017}).\\ \indent 
There exist fairly well-established experimental protocols to generate three-dimensional \he4 clusters of varying size \cite{Kuyanov2008,Toennies2022,Ulmer2024}; one could imagine adsorbing these clusters on a substrate to investigate their essentially 2D physics. 
With few exceptions \cite{Ketola1992,Ross1996} \he4 wets just about any surface; qualitatively different physical behavior can be expected, depending on the attractive strength and the corrugation of the substrate, which can vary considerably. On a weakly attractive substrate (e.g., lithium), whose corrugation can be neglected \cite{Hernandez2003,Boninsegni2023}, \he4 forms quasi-2D liquid-like superfluid puddles below equilibrium coverage \cite{Boninsegni1999,VanCleve2008}. On the other hand, on a strongly attractive, corrugated substrate such as graphite, a commensurate crystalline phase (known as the $C_{1/3}$) forms at the lowest coverage for which a stable full \he4 monolayer exists \cite{Bretz1973,Nielsen1980,Greywall1993}. Below that coverage, the system is predicted to feature coexistence of (macroscopic) solid clusters and vapor \cite{Pierce1999}. 
\\ \indent
On a substrate such as graphite, it is interesting to explore what happens as the size of a quasi-2D cluster is progressively reduced, down to the nanoscale limit. Superfluidity is expected to arise as a result of the competition of bulk and surface energetics, the latter frustrating the orderly arrangement of atoms promoted by the substrate corrugation. For intermediate size clusters, crystalline order and superfluidity may occur simultaneously, as predicted for parahydrogen clusters (in two and three dimensions) \cite{Mezzacapo2006,Mezzacapo2007,Mezzacapo09,Mezzacapo11,Gordillo2002}. Indeed, 2D clusters of parahydrogen comprising less than $\sim 25$ molecules behave like nanoscale ``supersolids'', with a high degree of molecular localization and the concomitant, frequent occurrence of quantum-mechanical molecular exchanges throughout the whole cluster, i.e., not restricted to a specific region thereof (e.g., its surface) \cite{Idowu2014}, and one may wonder if similar behavior could be observed in helium 2D clusters.
Of course, an important difference is that in parahydrogen clusters crystalline order originates from the interaction among molecules, whereas, in the case of helium, it is the corrugation of the underlying substrate that localizes atoms at lattice sites.  
\\ \indent
Experimentally, the stabilization of quasi-2D rare gas clusters has been demonstrated  by confining them  between two sheets of graphene \cite{Langle2024}. On this point, it is worth mentioning a recent microscopic theoretical study of quasi-2D superfluid \he4 clusters intercalated into graphite \cite{Ahn2020}. However, while several microscopic theoretical calculations for 3D \he4 clusters have been carried out, based on Quantum Monte Carlo (QMC) simulations and other methods \cite{Sindzingre1989,Whaley1994,Draeger2003,Dutra2017}, no comparable study has yet been reported for (quasi) 2D clusters. The inherent difficulty in simulating by QMC weakly bound clusters, which have a strong tendency to evaporate, is well understood; some workarounds exist, such as the use of an artificial external potential to keep the cluster together, but that clearly complicates the interpretation of the results. On the other hand, state-of-the-art QMC methodologies nowadays allow one to reach temperatures low enough to ensure cluster stability, thereby enabling reliable and accurate simulations. 
\\ \indent
In this paper, we present results of a QMC study of nanoscale size quasi-2D \he4 clusters adsorbed on a graphite substrate. Our computational methodology is the canonical variant \cite{Mezzacapo2006,Mezzacapo2007} of the continuous-space Worm Algorithm \cite{Boninsegni2006,Boninsegni2006b}, a robust technique allowing one to compute essentially exact numerical thermodynamic estimates for Bose systems at finite temperature.
We made use of an accepted microscopic model of the system, utilized in essentially {\em all} previous comparable studies. Specifically, we consider \he4 atoms as point-like particles of spin zero, moving in the presence of an infinite 2D substrate (graphite); the system is enclosed in a three-dimensional, rectangular simulation cell, the graphite substrate aligned with the basal $x$-$y$ plane; formally, periodic boundary conditions are utilized in all directions, but the size of the simulation cell (several tens of \AA\ in all directions) is much greater than that of the \he4 clusters, rendering boundary conditions unimportant. We model the interaction between two \he4 atoms using the standard Aziz pair potential \cite{Aziz1979}, while for the interaction of a \he4 atom with the substrate we utilize the Carlos-Cole potential \cite{Carlos1980}, which provides a quantitative description of the substrate corrugation. In order to assess the effect of corrugation, we have also carried out simulations using the laterally averaged version of this potential, i.e., assuming a flat substrate.
Technical details of the simulations are identical with those of Refs. \cite{Corboz2008,Happacher2013}. 
\\ \indent
Physical quantities of interest in this work, easily accessible in a QMC simulation based on the methodology utilized here, are the radial density profiles computed with respect to the center of mass of the cluster, providing cogent structural information, as well as the global superfluid response of the cluster, computed with the well-known ``area'' estimator \cite{Sindzingre1989}. We also made use of the local area estimator proposed in Ref. \cite{Kwon2006} to compute the radial superfluid fraction, again computed with respect to the center of mass of the cluster.
\begin{figure}
\includegraphics[width=\linewidth]{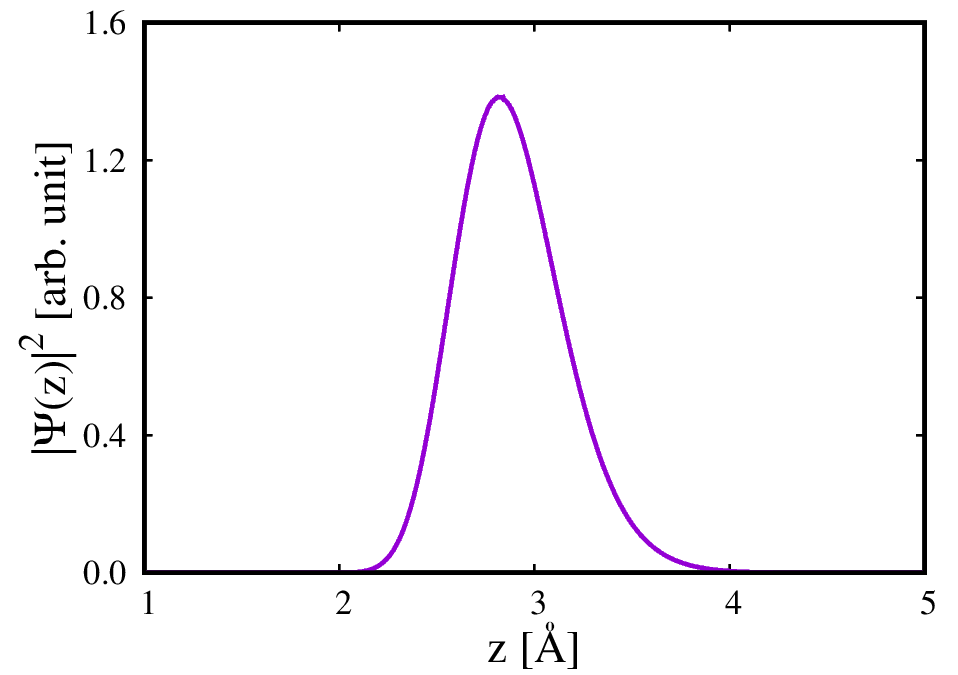}
\caption{Probability density of position (in arbitrary units) for $^4$He atoms, computed
with respect to the basal graphite plane, for a quasi-2D $^4$He cluster. The normalization is the same for all cluster sizes. Results for
clusters of different numbers of atoms cannot be distinguished, within the statistical uncertainties of our calculations, which are too small to be displayed.}
\label{nofz}
\end{figure}
\\ \indent
The results shown here are all obtained at a temperature $T=0.125$ K, which is low enough for us to draw the most important physical conclusions; we carried out simulations to other temperatures as well,  down to $T=0.03125$ K, and we shall comment on those as needed.
The first observation is that, regardless of whether the substrate is corrugated or smooth,  at sufficiently low temperature ($T\lesssim 0.15$ K) simulated clusters comprising as few as 7 $^4$He atoms stay together, i.e., no artificial confining potential is needed to prevent in-plane evaporation. While corrugation reduces in part the tendency to evaporate, by pinning atoms at preferred absorption sites, on a smooth substrate the system is held together (at sufficiently low temperature) by quantum-mechanical exchanges, which result in an effective attraction among identical bosons. 
\\ \indent
Fig. \ref{nofz} shows the probability density of position for $^4$He atoms in the direction perpendicular to the basal plane. The result shown in Fig. \ref{nofz} is independent of cluster size. The density profile is virtually indistinguishable from that of a full quasi-2D film adsorbed on graphite \cite{Corboz2008} or graphene \cite{Happacher2013,Yu2024}, regardless of whether one considers a corrugated or smooth substrate. In other words, there is no evidence of the kind of ``beading up'' in the perpendicular direction that takes place on weak substrates (such as Cs). Thus, this  system is a promising candidate for the experimental study of quasi-2D \he4 clusters.
\begin{figure}
\includegraphics[width=\linewidth]{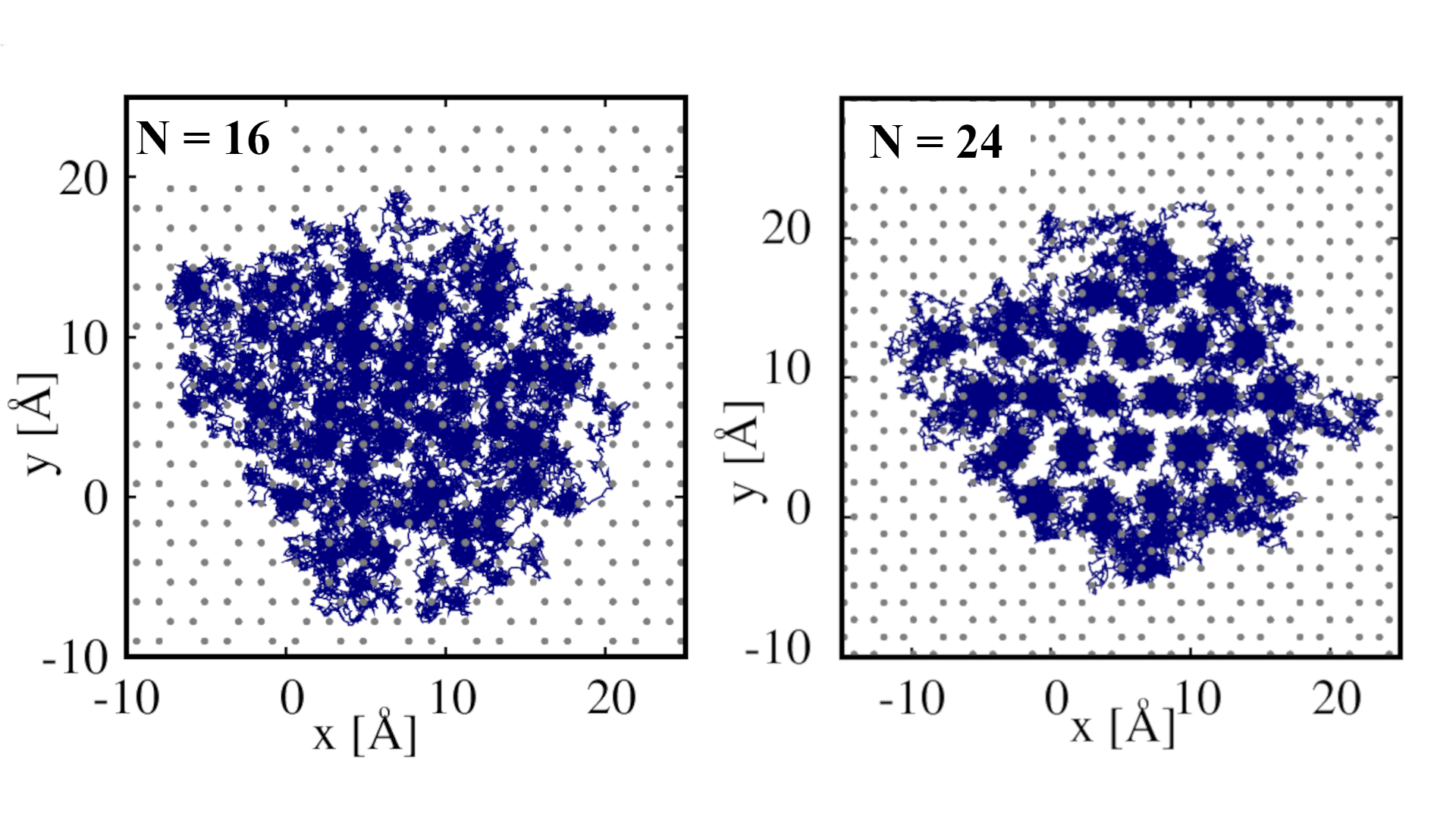}
\includegraphics[width=\linewidth]{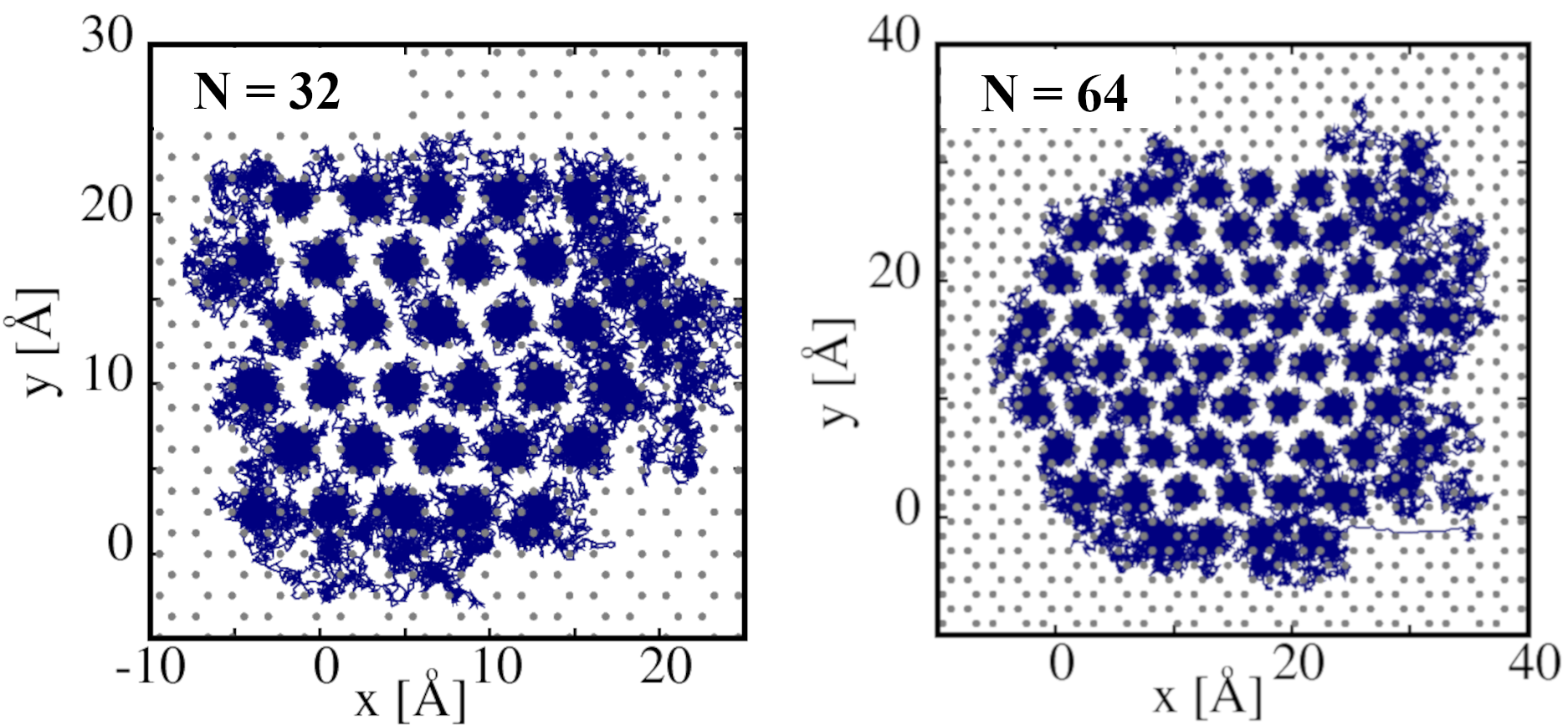}
\caption{Many-atom configuration snapshots (particle world lines projected on the graphite basal $x-y$ plane) for quasi-2D $^4$He clusters comprising 16, 24, 32 and 64 atoms. Dots arranged on a hexagonal lattice represent the carbon atoms on the graphite substrate.}
\label{snap12}
\end{figure}
\\ \indent
We now discuss the results obtained on a fully corrugated model of the substrate, contrasting them to those pertaining to a smooth substrate when the physical behaviors in the two cases differ substantially. 
Fig. \ref{snap12} shows representative instantaneous many-atom configurations (particle world lines), observed in the course of simulations, for clusters  comprising $N=16, 24, 32$ and 64 \he4 atoms. These images illustrate the qualitative structural change occurring when the number of atoms in the cluster exceeds $N\sim 20$. In the case of the smallest ($N=16$) cluster, no well-defined structure can be observed, and concurrently quantum-mechanical exchanges of indistinguishable \he4 atoms take place throughout the whole system. On the other hand, the greater clusters display incipient formation in their centers of the $C_{1/3}$ commensurate crystalline phase that characterizes a full monolayer; in this case, exchanges take place primarily at the surface of the cluster. This trend can be observed on the three clusters of size greater than $N=16$ shown in Fig. \ref{snap12}, with the central crystalline region growing in size, and only loosely bound atoms on the surface exchanging among themselves. 
Physically, this behavior can be interpreted as arising from the competition between surface and bulk energetics, which, for small cluster sizes, favors a disordered, liquid-like arrangement of atoms, with a crucial role being played by quantum exchanges \cite{Boninsegni2012b}.
\begin{figure}
\includegraphics[width=\linewidth]{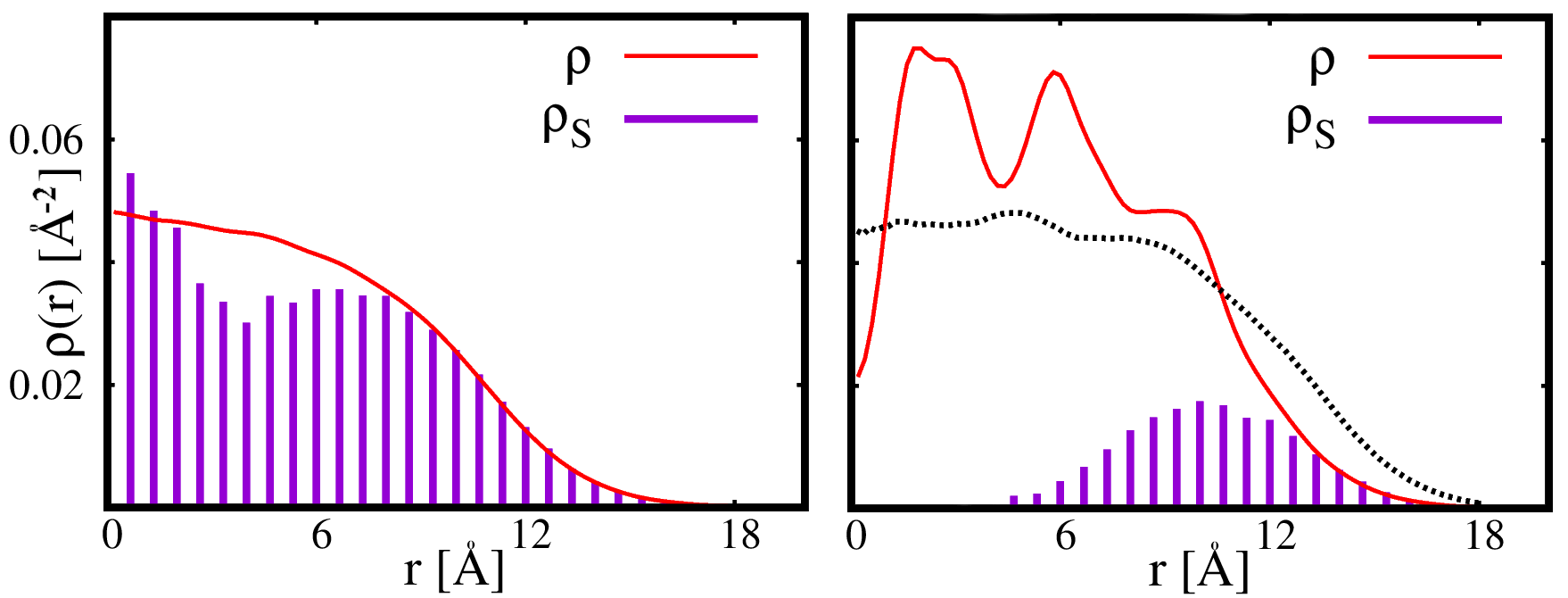}
\caption{Radial (solid line) 2D \he4 density $\rho$ for clusters comprising 16 (left) and 24 (right) atoms. Density profiles are computed with respect to the center of mass of the clusters. Spikes show the local superfluid density $\rho_S$. Dotted line shows the density profile computed on a smooth substrate. The apparently unphysical value of $\rho_S$ (greater than $\rho$) at short $r$ for the $N=16$ cluster results from the relatively large uncertainty affecting the determination of the local moment of inertia in the $r\to 0$ limit.}
\label{compare1624}
\end{figure}
\begin{figure}
\includegraphics[width=\linewidth]{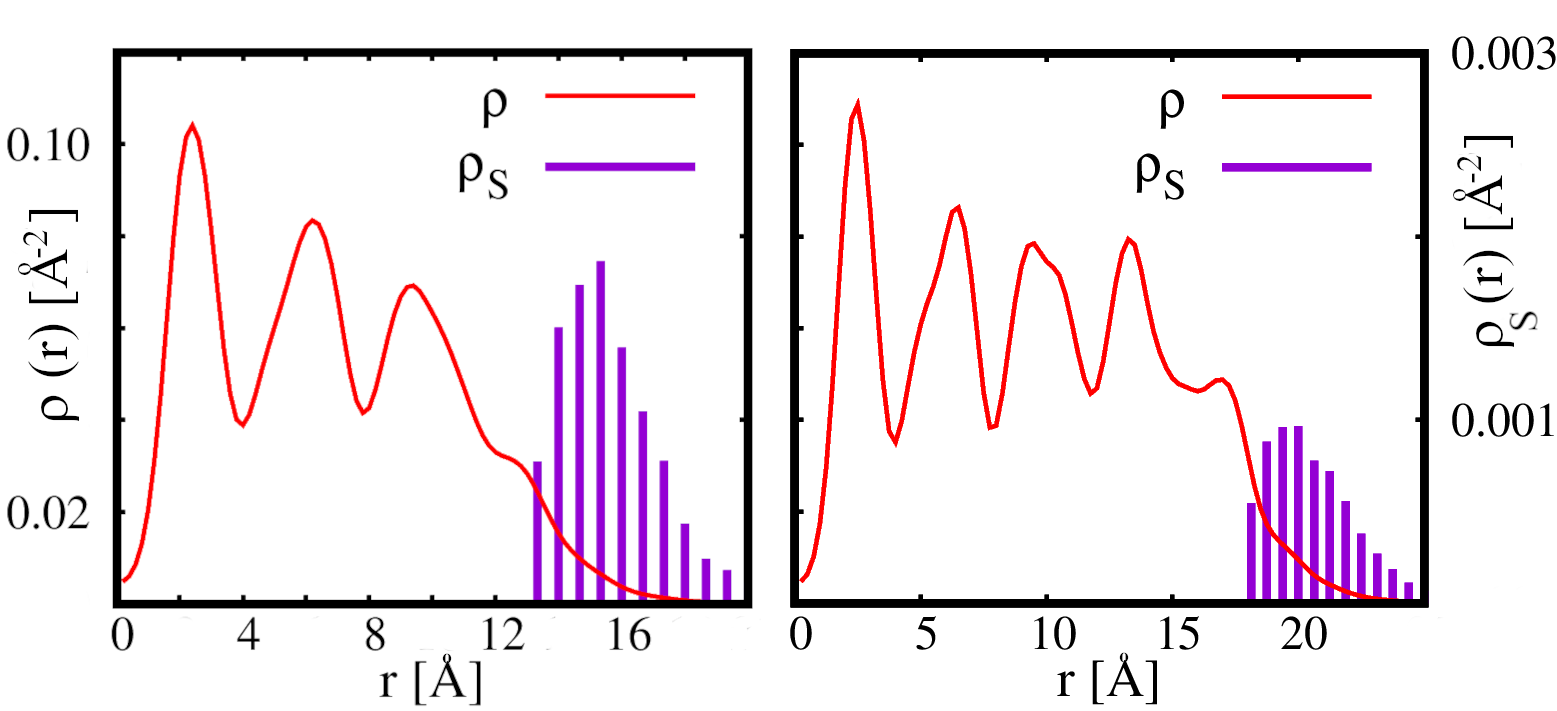}
\caption{Same as Fig. \ref{compare1624} but for $N=32$ (left) and $N=64$ (right). A different scale is used in this case for the superfluid density (right axis)}
\label{compare3264}
\end{figure} 
\\ \indent
These findings are {\em qualitatively} different from what is observed in 2D clusters of molecular parahydrogen, which can be regarded as nanoscale ``supersolids'', as they always  feature a well-defined, clearly identifiable solid-like structure (even with as few as 7 molecules), but with exchanges involving {\em all} the molecules in the clusters, at least for $N < 25$ (they become strongly suppressed for greater clusters) \cite{Idowu2014}. This underscores the strong pinning of \he4 atoms arising from the graphite substrate, as opposed to that ensuing from the mutual interaction of parahydrogen molecules.
\\ \indent
The evolution of the superfluid response of the \he4 clusters adsorbed on graphite, as the size of the cluster is increased, is the central aspect of this work. The global superfluid response, at the temperature considered here, is close to 100\% for clusters with fewer than 20 atoms. It decreases to around 50\% for $N=24$, and to less than 10\% for $N \ge 32$. Deeper insight is offered by the examination of the distribution of the superfluid response throughout the cluster, which is shown in Figs. \ref{compare1624} and \ref{compare3264}. It should be mentioned that, generally speaking, the statistical uncertainty affecting the quantities shown in the figures is fractionally small, typically barely detectable on the scales of the figures, as evidenced by the smoothness of the curves. The only exception is the local superfluid density, which (for a superfluid cluster) is affected by a large uncertainty in the center of the cluster (e.g., for $r \lesssim 3$ \AA\ for $N=16$).  \\ \indent
The radial density profile computed with respect to the center of mass of the cluster is featureless for the smallest ($N=16$) system shown, and essentially indistinguishable from that computed on a smooth substrate. Concurrently, the local superfluid response (shown by the spikes) is essentially uniform throughout the whole cluster, slightly suppressed only in the central part, where the cluster is fractionally ($\sim 75\%$) superfluid (as mentioned above, the calculation of the local superfluid fraction for small radial distances is affected by a large error, due to the uncertainty in the estimation of the local moment of inertia as $r\to 0$ \cite{Kwon2006}). We attribute this to the increased localization of the atoms in the center of the cluster, resulting from both the higher local density and the corrugation. The same physical picture is observed also for a cluster of $N=20$ \he4 atoms. Thus, the physics of clusters comprising twenty \he4 atoms or less is basically unaffected by the substrate corrugation.  
\\ \indent
On increasing the size of the cluster to $N=24$, the center begins to crystallize and the superfluid response is confined to the outer region, where quantum-mechanical exchanges persist. Indeed, as shown in Fig. \ref{snap12}, a clear separation can be drawn between the central area, where atoms are localized, and the outer region where instead the quantum-mechanical atomic ``clouds'' become fuzzier and overlap significantly. In this case, the radial density profile displays oscillations which reflect the orderly arrangement of atoms in the central part of the cluster, and is markedly different from that computed on a smooth substrate, as shown in the right panel of Fig. \ref{compare1624} (dotted line  is the profile computed on a smooth graphite substrate). Results obtained on a smooth substrate for all other clusters are very similar to that shown for $N=24$, i.e., the density profile is smooth, with no indication of shell formation, and the local superfluid density is uniform throughout the cluster.
\\ \indent
We carried out simulations of the $N=24$ cluster on the corrugated substrate down to a temperature as low as 30 mK, but saw no evidence of the {\em quantum melting} that takes place in small parahydrogen clusters, which are superfluid and liquid-like at low temperature but take on a crystalline structure as the temperature is increased \cite{Mezzacapo2006,Mezzacapo2007}. Rather, this cluster is seen to retain a solid-like structure at low temperature, with the radial density profile displaying little or no temperature dependence. At the same time, quantum exchanges are significantly enhanced and are not confined to the outer region but involve all of the atoms in the cluster. This behavior can be regarded as ``supersolid''; indeed, as indicated by the result shown in Fig. \ref{compare1624} (right panel), the superfluid response of this cluster is not confined to the outer, low density surface region, as seen in larger size clusters, but actually penetrates the more ordered, crystalline central part. Interestingly, this is qualitatively reminiscent of the physical behavior reported for three-dimensional \he4 clusters around a charged molecular impurity \cite{Marx2023}.
\\ \indent
All of this is a finite-size effect, however, as exchanges are rapidly suppressed with system size (they are essentially absent in the ground state of a commensurate crystalline monolayer \cite{Happacher2013,Yu2024}). Generally speaking, it is now accepted that no {\em bulk} supersolid phase  occurs in hard core, Lennard-Jones systems such as helium, but requires instead a ``soft core'' pairwise interaction among elementary constituents \cite{Boninsegni2012c,Kora2019}. 
Specifically, as the cluster size is increased the central crystalline region expands and superfluidity is progressively restricted to a thin surface layer, while atoms in the central region become increasingly localized as shown in Fig. \ref{compare3264}. The length of observed exchange cycles decreases sharply, and thus, concomitantly, the superfluid signal is  suppressed. For the largest cluster for which simulations were carried out, namely $N=128$, at temperature $T=0.125$ K the largest cycle of exchange observed in this work comprises only 7 atoms, i.e., a little over one tenth of all the \he4 atoms residing on the surface layer, with no measurable superfluid response at $T=0.125$ K.
\\ \indent
Altogether, the picture that emerges is that of a (quasi) one-dimensional  \he4 fluid, formed by loosely pinned \he4 atoms, residing in the outer region of an otherwise crystalline cluster. This quasi-1D system is expected to behave like a Luttinger liquid, analogously to what found in comparable physical settings for \he4 \cite{Boninsegni2001}.
On the other hand, if substrate corrugation is neglected, {\em all} of these clusters, including the largest one studied here (i.e., with $N=128$ \he4 atoms) are liquid-like and nearly 100\% superfluid at $T=0.125$ K. This reflects the propensity of helium to remain fluid, under the pressure of its own vapor, in the low temperature limit, in exception to otherwise universally observed crystallization. This remains true even in two dimensions, even in the presence of a strongly attractive potential like that of a smooth graphite substrate.
\\ \indent
In summary, we have carried out an extensive theoretical investigation, based on state-of-the-art QMC simulations, of nanoscale size clusters of \he4 adsorbed on graphite. We made use of the most realistic available microscopic model of the system, assessing quantitatively the effect of substrate corrugation.
Our main result is that quasi-2D clusters comprising as few as seven atoms are stable at low temperature ($T \lesssim 0.15$ K), with no evidence of in-plane evaporation. We find that sufficiently small clusters (less than 20 atoms) are liquid-like and superfluid, even in the presence of corrugation. Only when the size of the cluster exceeds more than approximately 20 atoms does the commensurate crystalline structure of a full monolayer begin to nucleate. \\ \indent 
These findings suggest a potential avenue to extend to two dimensions the investigation of superfluidity of finite systems, as well as the utilization of helium nanoclusters as an environment to perform quasi-2D molecular spectroscopy. On this point, we note that graphite may be a limiting case, as superfluidity is only predicted for relatively small clusters and that the substrate itself may be too strong for an impurity molecule to remain embedded into the cluster. On the other hand, superfluid  clusters of a greater size than what are observed on graphite are likely to be stable and observable on weaker substrates, such as those of alkali metals \cite{Boninsegni1999}, which are also less likely to affect the physical confinement of an impurity. These substrates may prove a more promising avenue for quasi-2D molecular spectroscopy.
\\ \indent 
There is evidence that the system investigated here, with a seemingly resilient outer layer of loosely bound $^4$He atoms, and a finite superfluid response that decreases with cluster size, might provide a realization of a quasi-1D  (Luttinger) liquid, or even of a Transverse Quantum Fluid (TQF), which is underlain by ``superclimbing'' modes \cite{Zhang2024}. Offering an authoritative, definitive answer requires the calculation of spatial correlation functions and  finite-size scaling analysis for systems of significantly greater size than that accessible in this investigation. 
\\ \indent
M. B. acknowledges support from the Natural Sciences and Engineering Research Council of Canada.

\bibliography{references}
\end{document}